\documentclass[preprint,eqsecnum,aps,showpacs,floatfix]{revtex4}
\usepackage{dcolumn} \usepackage{graphicx} \usepackage{amsmath}
\usepackage{amssymb} \usepackage{epsfig} \usepackage{float}
\usepackage{latexsym} \usepackage{rotating}
\usepackage{subfigure}
\usepackage{color}
\usepackage{psfrag}

\newcommand{\sgn}{\text{sgn}}

    \DeclareMathOperator{\sech}{sech}
\begin{document}
\title{Geodesic congruences in warped spacetimes}

\author{Suman Ghosh $^{1}$\footnote{E-mail:suman@cts.iitkgp.ernet.in},
Anirvan Dasgupta $^{2,3}$ \footnote{E-mail:anir@cts.iitkgp.ernet.in} and 
Sayan Kar$^{1,3}$ \footnote{E-mail:sayan@cts.iitkgp.ernet.in} }
\affiliation{$^1$Department of Physics and Meteorology, Indian Institute of Technology, Kharagpur 721 302, India \\
$^2$Department of Mechanical Engineering, Indian Institute of Technology, Kharagpur 721 302, India \\
$^3$Centre for Theoretical Studies, Indian Institute of Technology, Kharagpur 721 302, India}


\begin{abstract}
In this article, we explore the kinematics of timelike geodesic congruences 
in warped five dimensional bulk spacetimes, with and without thick or thin 
branes. Beginning with geodesic 
flows in the Randall--Sundrum AdS (Anti de Sitter) geometry without and with 
branes we find analytical expressions for the expansion scalar and 
comment on the effects of including thin branes on its evolution. 
Later, we 
move on to congruences in
more general warped bulk geometries with a cosmological 
thick brane and a time-dependent extra dimensional scale. 
Using analytical expressions for the velocity field, 
we interpret the expansion, shear and rotation (ESR) along 
the flows, as functions of the extra dimensional coordinate. 
The evolution of a cross-sectional area orthogonal to the congruence, 
as seen from a local observer's point of view, is also shown graphically. 
Finally, the Raychaudhuri and geodesic equations in backgrounds with a 
thick brane are solved numerically in order to figure out  
the role of initial conditions (prescribed on the ESR) and 
spacetime curvature on the evolution of the ESR.

\end{abstract}

\maketitle

\section{Introduction}

Almost a century ago, in their pioneering research \cite{kk}, Kaluza and Klein 
(KK) proposed unification of four dimensional gravity and electromagnetism 
in a five dimensional gravity framework. This proposal raised a fair amount 
of curiosity, interest and research activity among theoretical physicists, 
on the physics of extra dimensions. The idea of extra spatial dimensions appeared in a new incarnation with the advent of Superstring theories \cite{string,Antoniadis:1990ew}. 
The theoretical existence of branes in String theory eventually motivated 
the hypothesis that we may be living on an embedded, timelike submanifold 
(the brane) of a higher dimensional ($D>4$) Lorentzian spacetime 
(warped or unwarped), as assumed in the  so-called Arkani-Hamed--Dvali--Dimopoulos (ADD) \cite{add,Antoniadis:1998ig} and Randall--Sundrum (RS) \cite{rs1} braneworld models.

The seminal work of Randall and Sundrum (RS) \cite{rs1,rs2} on warped 
braneworlds, published more than a decade ago, refers to the idea of the 
scale of the 
extra dimension being spacetime dependent, while addressing the issue of 
stability, in a two-brane scenario. In a single brane scenario or from a 
purely higher dimensional bulk perspective, the space-time dependence of 
the metric function(s) associated with the extra dimensional coordinate(s) 
basically imply that the scale of the extra dimension depends on the on-brane 
(four dimensional) spacetime coordinates. 
Except for a brief discussion on RS type models, we 
shall, in this article, mostly work with the single brane scenario 
and a five dimensional bulk.

In earlier papers \cite{Ghosh:2009ig, seahra}, geodesics in warped 
spacetimes have been investigated in detail. However, such a study of 
geodesics alone cannot tell us about the overall local behavior of a family 
of test particles, as observed in the neighbourhood of a freely falling 
observer. This motivates us to study the evolution of geodesic congruences.
Since the appearance of {\em Raychaudhuri equations}, in 1955 \cite{ray}, 
relativists have discussed and analysed its implications in various contexts. 
In its original incarnation, the Raychaudhuri equations provided the basis 
for the analysis of spacetime singularities in gravitation 
and cosmology \cite{hawk}. For example, the equation for the expansion  and 
the resulting theorem on geodesic focusing, is a crucial ingredient in the 
proofs of Penrose-Hawking singularity theorems \cite{Penro,Haw}.

The kinematics of geodesic congruences is characterised by three 
kinematical quantities: isotropic expansion, shear and rotation (henceforth referred as ESR) \cite{toolkit,wald,joshi,ellis,ciufolani,review}, which evolve 
along the flow according to the Raychaudhuri equations. Though mostly quoted 
and used in the context of gravity, these equations by virtue of their geometric nature, have a much wider scope in studying geodesic as well as 
non-geodesic flows in nature, which may possibly arise in diverse contexts 
(see \cite{review} for some open issues). Two of the authors here have 
recently used these equations to investigate the kinematics of geodesic 
flows in stringy black hole spacetimes \cite{adg3} and flows on flat and 
curved deformable media (including elastic and viscoelastic media) in detail 
\cite{adg1,adg2}. 

In this article, we attempt to understand the kinematics of geodesic flows in five dimensional warped bulk spacetimes with and without branes. 
In Section II we quickly recall the background spacetime geometries 
and geodesics. Section IIIA analyses flows in the RS geometry with and
without branes. The ESR, as obtained from definitions, for a background with
a thick brane are discussed in Section IIIB. Numerical solutions of the
geodesic and Raychaudhuri equations are presented in IIIC. Finally, Section
IV contains our conclusions and comments.

\section{The bulk spacetimes and geodesics}



The bulk spacetimes we work with are given by the line element 
\cite{Ghosh:2009ig} ($\eta$ is the conformal time),
\begin{equation}
ds^2 = e^{2f(\sigma)} a^2(\eta) [- d\eta^2 + d{\bf X}^2] + b^2(\eta) d\sigma^2 .  \label{eq:cmetric}
\end{equation}
Table \ref{tab:models}, shows the chosen functional forms of $f(\sigma)$ 
(the warp 
factor), the cosmological [$a(\eta)$] and extra dimensional [$b(\eta)$] scale 
factors (following \cite{Ghosh:2009ig}). 
\begin{table}[h]
\begin{center}
\begin{tabular}{|c|c|}
\hline
$f(\sigma)$ & \{  $a(\eta)$, $b(\eta)$ \} \\
\hline
 $ -\ln\,(\cosh\, k\sigma)$ & Set (A)\,\, \{$ 2\eta,1 + \frac {1}{\eta}$ \} \\
Decaying warp factor & FRW (Radiation dominated) brane\\
\hline
 $ \ln\,(\cosh\, k\sigma)$& Set (B) \,\, \{ $\frac {1} {1 - \eta}, 1 - \frac{\eta}{2}$  \} \\
Growing warp factor & de Sitter brane\\
\hline
\end{tabular}
\end{center}
\caption{The four possible combinations of $f(\sigma)$, $a(\eta)$ and $b(\eta)$.}
\label{tab:models}
\end{table}
We mostly prefer to work with thick branes in order to avoid the discontinuities and delta functions which appear in the connection and curvature, for thin branes. However, as we will see later, in some special cases (e.g. Einstein spaces) one can indeed solve the Raychaudhuri equations consistently, with zero rotation and shear, in the presence of thin branes. It may be noted that these models are assumed to represent the evolution of the universe beginning at a finite time when both the scales of visible and extra dimension were same.
The geodesic equations 
in the spacetime (\ref{eq:cmetric}) are almost impossible to solve analytically. However, they may be recast as a dynamical system of coupled, ordinary, first order differential equations as given by Eqs 3.8-3.11 in \cite{Ghosh:2009ig}.
Thus, for simple cases, knowing the first integrals one may directly 
determine the ESR.

\section{Raychaudhuri Equation and ESR variables}

\subsection{Randall--Sundrum warp factor with and without branes}

\begin{figure}[h!]
\subfigure[
]{\label{fig:domain1}\includegraphics[width = 2.5 in]{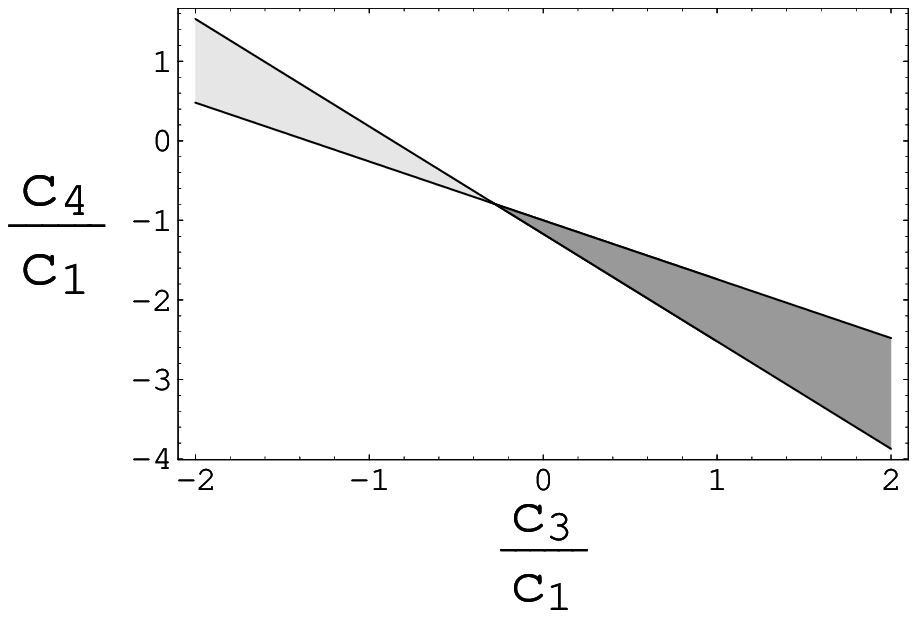}}\hspace{0.5 in}
\subfigure[]{\label{fig:2brane2}\includegraphics[width = 2.5 in]{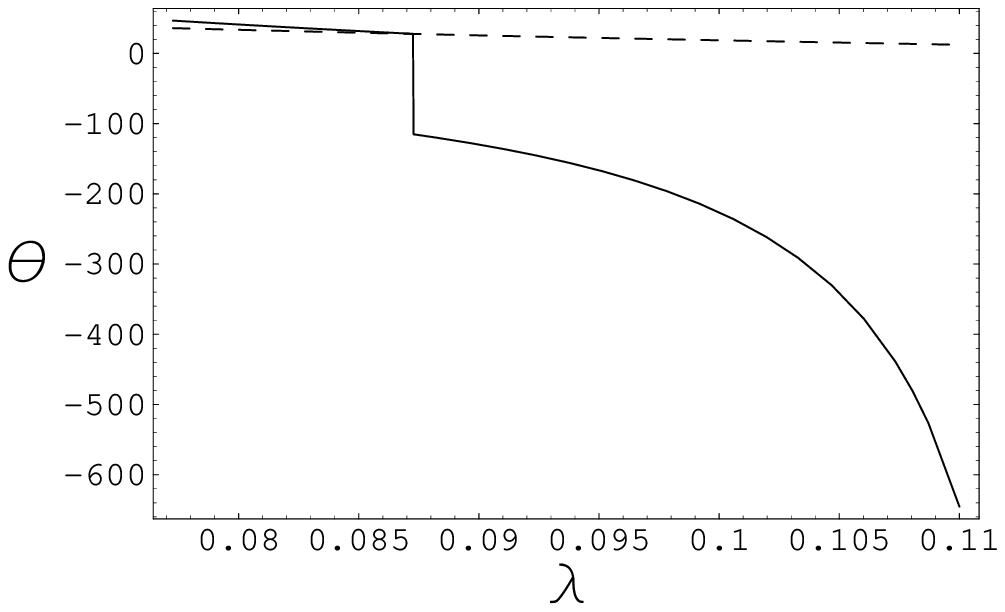}}
\caption{
(a) Phase diagram in the $c_3/c_1$-$c_4/c_1$ plane. Points lying in the deeply (lightly) shaded region correspond to focusing (defocusing) of geodesics.
(b) Evolution of expansion scalar in presence (continuous line) and in absence (dotted line) of two 3-branes with $c_1 = 1$, $c_3 = 1$ and $c_4 = -2$ which lies in the dark shaded region in (a).
Specific values of the parameters are: $k = 12$, $C = 2$, $\lambda_0 = 0.087$ (1st brane location),  $\lambda_{\pi} = 0.131$ (2nd brane location),  $c_2 = 0$ and $\alpha = 6.9282$ (from Eq.\ref{eq:transcend1}).} \label{fig:2brane}
\end{figure}

In Einstein spaces, $R_{AB} \sim g_{AB}$ and 
the equation for the expansion simplifies considerably if we 
assume the shear and rotation as zero. 
For the RSI scenario \cite{rs1,rs2} in the absence of any brane, the Raychaudhuri equation \cite{wald,toolkit} gives us
\begin{equation}
\frac{d\Theta}{d\lambda} +\frac{\Theta^2}{4} = \frac{\Lambda}{6M^3} \label{eq:ray1_1}
\end{equation}
where, $\Lambda$ is the bulk cosmological constant and $M$ is the five 
dimensional Planck mass. With $\Theta = 4 \frac{\dot F}{F}$ (notion of focusing is related to $F = 0$, $\dot F < 0$ at finite $\lambda$), leads to
\begin{equation}
\ddot F + k^2 \, F = 0, \hspace{0.5cm} \mbox{with} \hspace{0.5cm} k = \sqrt{\frac{-\Lambda}{24M^3}}\label{eq:ray1_3}, 
\end{equation} 
Following \cite{rs2}, $\Lambda$ is negative and 
Eq.\ref{eq:ray1_3} has simple oscillatory solutions 
such as $c_1 \sin(k\lambda + c_2)$, which imply $\Theta = 4k\,\cot(k\lambda + c_2)$. Therefore, the nature of focusing or defocusing of geodesics in the bulk 
depends on the initial condition or the value of $c_2$. However, 
this is the behavior of geodesic congruences in the bulk with no branes. 
If we introduce two 3-branes, the hidden brane (with positive tension $24M^3k$) at $\sigma = 0$ and the visible brane (with equal negative tension) at $\sigma = \pi$, Eq.\ref{eq:ray1_3} becomes
\begin{equation}
\ddot F + \left[k^2 + 2k \{\delta(\sigma) - \delta(\sigma - \pi)\}\right] \, F = 0 \label{eq:ray1_4}.
\end{equation}  
 Using the following property of Dirac delta function
\begin{equation}
\delta(\sigma(\lambda)) =  \sum_i \frac{\delta(\lambda - \lambda_i)}{|\dot \sigma(\lambda_i)|} \label{eq:delta1},
\end{equation}  
and the first integral of the $\sigma$ geodesic equation
\begin{equation}
\dot\sigma = \sqrt{C^2 e^{-2f} - 1}  \hspace{0.5cm} \mbox{with} \hspace{0.5cm} f(\sigma) = - k|\sigma|,
\end{equation}  
we arrive at
\begin{equation}
\ddot F + \left[k^2 + k_1 \delta(\lambda - \lambda_0) -  k_2 \delta(\lambda - \lambda_\pi) \right] \, F = 0 , \label{eq:ray1_5}
\end{equation}  
where, $k_1 = \frac{2k}{\sqrt{C^2 - 1}}$, $k_2 = \frac{2k}{\sqrt{C^2e^{2k\pi} - 1}}$, $\lambda_0 = \frac{\sec^{-1}C}{k}$ and $\lambda_\pi = \frac{\tan^{-1}\sqrt{C^2e^{2k\pi} - 1}}{k}$. The solution of the above equation is given by \cite{doubledelta}
\begin{equation}
F(\lambda) = c_1 \sin(k\lambda + c_2) + c_3 e^{-\alpha |\lambda - \lambda_0|} + c_4 e^{-\alpha |\lambda - \lambda_\pi|} \label{eq:2ndorder}
\end{equation}  
where $c_1$, $c_2$, $c_3$, $c_4$ are arbitrary constants. 
We then integrate the second order equation for F, around the neighborhood
of $\lambda_0$ and $\lambda_\pi$ to obtain two algebraic equations for
$c_1$ and $c_2$. The determinant condition for nontrivial solutions of 
$c_1$, $c_2$ yields the following transcendental equation,
\begin{equation}
(2\alpha - k_1)(2\alpha + k_2) + k_1 k_2 e^{-2\alpha (\lambda_\pi - \lambda_0)} = 0 \label{eq:transcend1}.
\end{equation}  
from which $\alpha$ can be obtained numerically.
In our case, $\alpha \sim k_1/2$ is a good approximation
(we have checked this with numerical solutions as well). 
Fig.\ref{fig:domain1} shows (shaded regions) the domain of 
the parameters $c_3/c_1$ and $c_4/c_1$ ($c_2$ is taken to be zero), 
in cases where $F = 0$ at some finite value of $\lambda$ between the 
two branes. Points on the boundaries of the shaded regions correspond to the occurrence of $F = 0$ at the location of the branes
(i.e. $\lambda=\lambda_0, \lambda=\lambda_\pi$ and the corresponding $\sigma$
through the function $\sigma(\lambda)$). 
In the lightly shaded region, $\dot F > 0$, which implies complete defocusing 
of geodesics ($\theta\rightarrow +\infty$) at finite $\lambda$, 
whereas in the darker region, $\dot F < 0$, i.e. 
geodesic focusing ($\theta\rightarrow -\infty$) is possible. Note that 
different values for $c_2$ will result in different parameter space diagrams 
involving the quantities $c_3/c_1$ and $c_4/c_1$.
Eq.\ref{eq:2ndorder} leads to the modified expansion scalar, due to the presence of the branes, as given by
\begin{equation}
\Theta(\lambda) = 4\frac{ c_1k \cos(k\lambda + c_2) - c_3 \alpha\, \sgn(\lambda - \lambda_0) e^{-\alpha |\lambda - \lambda_0|}  - c_4 \alpha\, \sgn(\lambda - \lambda_\pi) e^{-\alpha |\lambda - \lambda_\pi|} }{ c_1 \sin(k\lambda + c_2) + c_3 e^{-\alpha |\lambda - \lambda_0|} + c_4 e^{-\alpha |\lambda - \lambda_\pi|} }. \label{eq:theta_rs1}
\end{equation}  
Due to the new integration constants, the behavior of geodesic 
congruences have become much richer. One such scenario is shown in Fig.\ref{fig:2brane2}, where for specific values of these parameters, $c_1 = 1$, $c_3 = 1$ and $c_4 = -2$ which belongs to the dark shaded domain in Fig.\ref{fig:domain1}, gives rise to occurrence of congruence singularity in between the brane locations.

\subsection{Calculating ESR from velocity field}

One may derive analytic expressions for the ESR variables directly from 
the following definitions \cite{wald,toolkit} for the expansion $\Theta$, the shear $\Sigma_{AB}$ and the rotation $\Omega_{AB}$, using the geodesic vector field components obtained in \cite{Ghosh:2009ig}, 
\begin{eqnarray}
\Theta &=& \nabla_A u^A  \label{eq:thetadef},\\
\Sigma_{AB} &=& \frac{1}{2} \left (\nabla_B u_A +\nabla_A u_B\right ) -\frac{1}{n-1} h_{AB}\Theta \label{eq:sigmadef}, \\
\Omega_{AB}&=&\frac{1}{2} \left ( \nabla_B u_A - \nabla_A u_B\right ).\label{eq:omegadef}
\end{eqnarray}
Here, $n$ is the dimension of spacetime and $h_{AB} = g_{AB} \pm u_A u_B$ is the projection tensor (the plus sign is for timelike curves whereas the minus one is for spacelike ones) and $u_A u^A =\mp 1$. 
Let us now obtain the ESR for some specific cases (for thick branes), 
where we take one or 
two of the metric functions as constants. 

\subsubsection{Case 1: $a(\eta) = b(\eta) =$ constant} \label{case1}

Here, with only a non--constant $f(\sigma)$ present in \ref{eq:cmetric}, 
the velocity field components for timelike geodesics are
\begin{eqnarray}
u^{\alpha} &=& C_{\alpha} e^{-2f} \hspace{1cm} \mbox{where}\hspace{0.5cm}\alpha = 0,1,2,3 \\
u^4 &=& \sqrt{C^2 e^{-2f} - 1} \hspace{1cm} \mbox{where}\hspace{0.5cm} C^2 = C_0^2 - \sum_{i=1}^3C_i^2,
\end{eqnarray}
where the $C_{\alpha}$'s are integration constants, which are constrained by the fact that $u^4$ has to be real valued. 

According to the definitions Eq.\ref{eq:thetadef} - Eq.\ref{eq:omegadef}, the expansion scalar and the other ESR variables for a congruence of timelike geodesics, written as functions of $\sigma$, give us,
\begin{equation}
\Theta = \frac{3C^2e^{-2f} - 4}{\sqrt{C^2 e^{-2f} - 1}} f',
\end{equation}
\begin{equation}
\Sigma^2 = \Sigma_{AB}\Sigma^{AB}  = \frac{3C^4 e^{-4f} f'^2}{4(C^2e^{-2f} - 1)}\hspace{0.5cm} \mbox{and} \hspace{0.5cm} \Omega_{AB} = 0 \, \forall \, A,  B.    \label{eq:srparameters}
\end{equation}
For the chosen growing and decaying warp factors respectively, the expansion scalar becomes 
\begin{equation}
\Theta_+ = \frac{3C^2 \sech^2\sigma - 4}{\sqrt{C^2 \sech^2\sigma - 1}} \tanh\sigma \hspace{.5 cm}\mbox{and}\hspace{.5 cm}
\Theta_- = - \frac{3C^2 \cosh^2\sigma - 4}{\sqrt{C^2 \cosh^2\sigma - 1}} \tanh\sigma.
\end{equation}
It can be easily seen that, for $\dot\sigma > 0$ and $C>1$, $\Theta_{\pm} \rightarrow -\infty$ as $\sigma$ increases. Therefore a geodesic congruence singularity arises in both the cases. With $\Theta_+$, if $C > \sqrt{4/3}$, initially the expansion remains positive but eventually geodesics meet exactly at the boundary of the accessible domain along the extra dimension because the velocity component along $\sigma$, $u^4$, which appears in the denominator in the expression for $\Theta$, vanishes at that point (and also changes sign). With $\Theta_-$, the geodesic congruence singularity appears as $\sigma \rightarrow \infty$. For $\dot\sigma < 0$ an exactly similar behavior is obtained. 
From the nature of corresponding geodesics (see \cite{Ghosh:2009ig}), we note that $\Theta_+$ experiences a finite time singularity (i.e. finite $\lambda$ as well as finite $\sigma$) but $\Theta_-$ becomes singular at finite $\lambda$ but $\eta,\sigma \rightarrow \infty$. 

\begin{figure}[h]
\subfigure[\,Evolution of the projected square element on $\sqrt{g_{11}}\xi^1$-$\sqrt{g_{22}}\xi^2$ plane]{\label{fig:surf1_1}\includegraphics[scale=1.2]{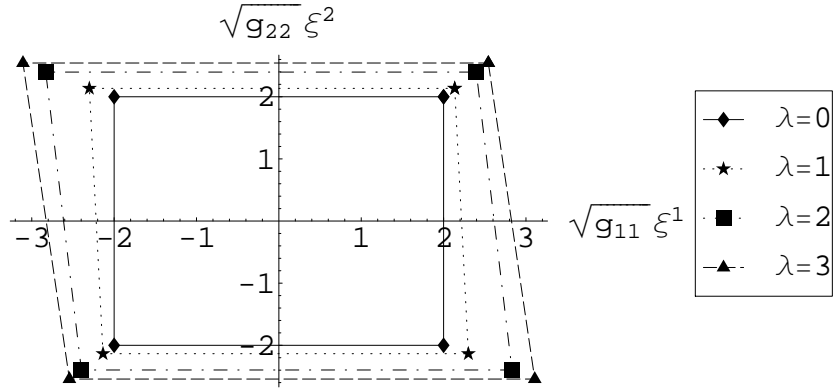}} 
\subfigure[\,Evolution of the projected square element on $\sqrt{g_{11}}\xi^1$-$\sqrt{g_{44}}\xi^4$ plane]{\label{fig:surf1_2}\includegraphics[scale=1.2]{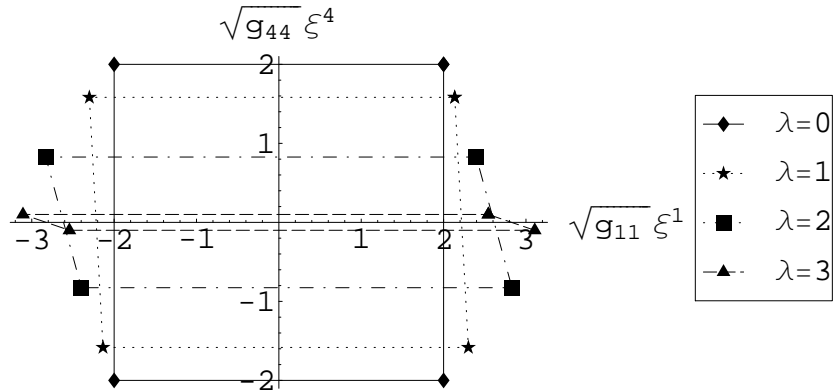}}
\caption{Evolution of different 2D surface elements in Case 1 in presence of a growing warp factor} \label{fig:surf1}
\end{figure}

To understand how geodesic congruences behave in all the abovementioned scenarios, let us consider the evolution of the projections of the cross-sectional area orthogonal to the flow lines, of a congruence of four geodesics, on different two dimensional surfaces. This is done by numerically solving 
\begin{equation}
u^C\nabla_C B_{AB} = - B_{AC}B^C_{\,B} - R_{ACBD} u^Cu^D \label{eq:gen}.
\end{equation}
\begin{equation}
\xi^{A}_{\,\,;B}u^B = B^A_{\,\,B} \xi^B \label{eq:xieq}
\end{equation}
along with the geodesic equations.
Here, $B_{AB}= \nabla_B u_A$ is the gradient of velocity field and $\xi^A$ represents the separation between two neighboring geodesics. Eq.\ref{eq:xieq} is essentially the evolution equation for the deviation vector. To see the evolution from a local observer's viewpoint, we have to express 
the tensorial quantities in the frame basis. The metric tensor in coordinate 
basis and frame basis are related as
\begin{equation}
g^{AB} = e^A_{\,\,a}\,e^B_{\,\,b}\, \eta^{ab},
\end{equation}
where the vierbein field, $e^A_a$, has two indices, ``$A$'' labels the general spacetime coordinate (w.r.t. the coordinate basis) and ``$a$'' labels the local Lorentz spacetime or local laboratory coordinates (w.r.t. the frame basis). The tensorial components in these two bases are related as
\begin{equation}
\xi^A = e^A_{\,\,a}\,\xi^a.
\end{equation}
 In the frame basis, we set the initial conditions such that, at $\lambda = 0$, the projected area has the shape of a square in the $\sqrt{g_{11}}\xi^1$-$\sqrt{g_{22}}\xi^2$ plane or in the $\sqrt{g_{11}}\xi^1$-$\sqrt{g_{44}}\xi^4$ plane where $\xi$'s are essentially solutions of Eq.\ref{eq:xieq}. Fig.\ref{fig:surf1} shows the evolution of the area elements as $\lambda$ increases. The origin represents the location of the observer. We have chosen initial conditions such that all components of the rotation vanish.

Fig.\ref{fig:surf1} shows how a projected 2D square element evolves, in a static bulk, in presence of a growing warp factor. In Fig.\ref{fig:surf1_1}, the initial square area (at $\lambda = 0$) in the $\sqrt{g_{11}}\xi^1$-$\sqrt{g_{22}}\xi^2$ plane expands and distorts slightly and converges on a parallelogram at $\lambda \sim 3.14$. In Fig.\ref{fig:surf1_2}, however the area in the $\sqrt{g_{11}}\xi^1$-$\sqrt{g_{44}}\xi^4$ plane shrinks, distorts and converges on the $\sqrt{g_{11}}\xi^1$ axis at $\lambda \sim 3.14$, clearly suggesting focusing along the extra dimension. 
The effect of shear is evident from the evolution of the shape of the area. 

\begin{figure}[h!]
\subfigure[\,Evolution of the projected square element on $\sqrt{g_{11}}\xi^1$-$\sqrt{g_{22}}\xi^2$ plane]{\label{fig:surf2_1}\includegraphics[scale=1]{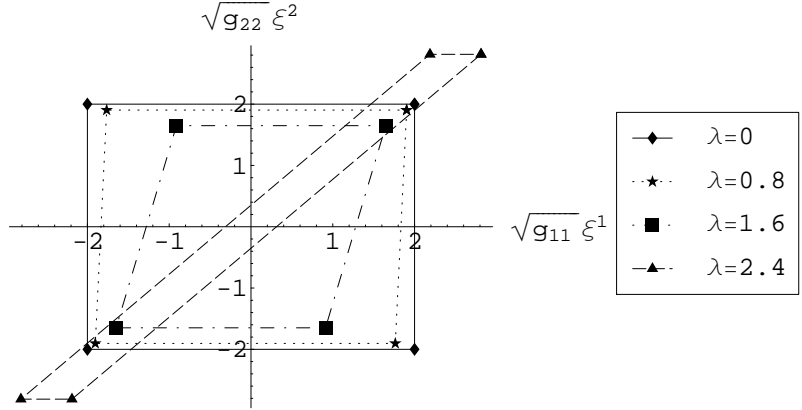}} 
\subfigure[\,Evolution of the projected square element on $\sqrt{g_{11}}\xi^1$-$\sqrt{g_{44}}\xi^4$ plane]{\label{fig:surf2_2}\includegraphics[scale=1]{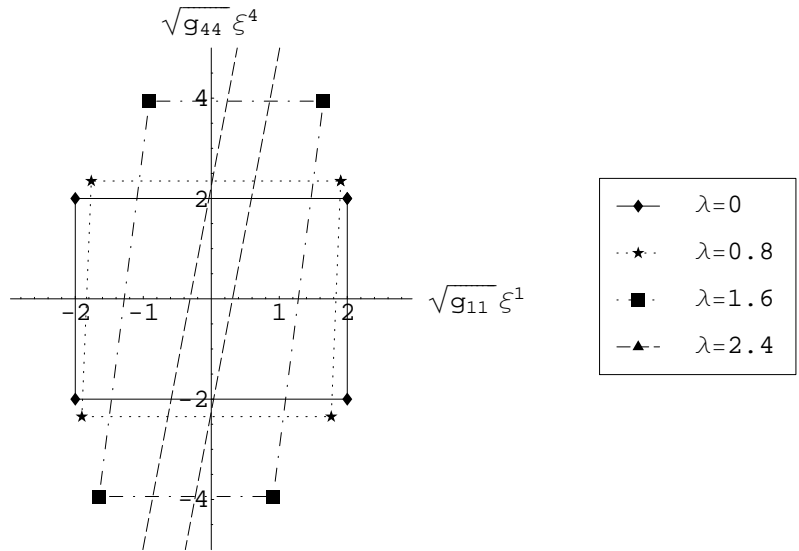}}
\caption{Evolution of different 2D surface elements in Case 1 in presence of a decaying warp factor.} \label{fig:surf2}
\end{figure}
On the other hand, Fig.\ref{fig:surf2} corresponds to the scenario where the warp factor is of decaying type. 
In Fig.\ref{fig:surf2_1}, shrinking of the area element is quite prominent 
whereas in Fig.\ref{fig:surf2_2} it is not so. However, in both the figures, 
the square area becomes more and more parallelogram shaped and eventually 
converge on a line $\sqrt{g_{11}}\xi^1\propto\sqrt{g_{22}}\xi^2$ or $\sqrt{g_{11}}\xi^1\propto\sqrt{g_{44}}\xi^4$ as $\lambda\rightarrow\infty$. 

The nature of evolution of the square elements is therefore a 
distinguishing feature between bulk universes with growing and decaying warp 
factors. It is worth mentioning here that, to a brane based observer, only the 
evolution depicted in Fig.\ref{fig:surf1_1} or Fig.\ref{fig:surf2_1} will be 
visible whereas congruence singularities realised in Fig.\ref{fig:surf1_2} or Fig.\ref{fig:surf2_2} will remain unnoticed. In a way, therefore, one can
find the nature of warping as well as the existence of extra dimensions 
from the evolution pattern of cross--sectional area elements.      

\subsubsection{Case 2: $f(\sigma)$=constant}\label{case2}

In this case,  with only non--constant $a(\eta),\;b(\eta)$ present in \ref{eq:cmetric}, 
the velocity field components for timelike geodesics are
\begin{eqnarray}
u^i &=& \frac{C_i}{a^2}, \hspace{0.75cm} \mbox{}\hspace{0.75cm} u^4 = \frac{C_4}{b^2}, \nonumber\\
u^0 &=& \sqrt{\frac{1}{a^2} + \sum_{i=1}^3 \frac{C_i^2}{a^4} + \frac{C_4^2}{a^2b^2}} 
\end{eqnarray}
We can calculate the expansion scalar which turns out to be,
\begin{equation}
\Theta = -\frac{1}{u_0} \left[3\frac{\dot a}{a} + 2\sum_i \frac{C_i^2\dot a}{a^3} + \frac{3C_4^2\dot a}{ab^2} + \frac{\dot b}{b} + \sum_i \frac{C_i^2\dot b}{a^2b} \right]. \label{eq:theta2}
\end{equation}
As done before one can also find the components of the shear tensor 
(not shown here). The rotation tensor components are zero, as is evident 
from the velocity field. 

In case of Set(A) [as given in Table \ref{tab:models}], the geodesics become parallel as $\eta \rightarrow \infty$. This is because, with a FRW (radiation dominated) brane, the expansion of the universe itself slows down with increasing $\eta$ (it is worth mentioning here that as the cosmological evolution is assumed to begin at a finite $\eta$, the past singularity will not appear here since it corresponds to $\eta = 0$ and falls outside the domain of $\eta$ for the models considered in this article).
For Set(B) [see Table \ref{tab:models}], the geodesics spread apart at an ever increasing rate with increasing $\eta$ 
-- this is due to very rapid (exponential in real time) expansion of the 
universe. In both these cases 
geodesic focusing is not achieved. On the other hand, as noted in 
the previous subsection, geodesic singularities are unavoidable when 
a non--constant warp factor is assumed. 
Therefore, it should be interesting to see how these two apparently opposite 
features compete with each other when we consider the full braneworld geometry 
with all three non--constant metric functions present.

We end this subsection by trying to figure out the individual effect of the 
dynamic nature of $b(\eta)$. Assuming $a(\eta)$ as constant in Eq.\ref{eq:theta2} we get,
\begin{equation}
\Theta = \frac{\left(1 + \sum_iC_i^2 \right) \dot b}{\sqrt{\left(1 + \sum_iC_i^2\right)b^2 + C_4^2}} .\label{eq:theta2a}
\end{equation}
In our models $\dot b$ is negative and as $\eta$ increases it tends to zero. 
For the $b(\eta)$ of Set(A), $\Theta$ tends to zero i.e. geodesics become 
parallel while for Set(B) it converges to a finite value which implies that 
the geodesics keep  moving away from each other at an approximately steady rate.So, with only a $b(\eta)$, in both the above Sets (A) and (B), 
geodesic focusing never happens. It is interesting to note that even if 
$b(\eta) \rightarrow 0$ i.e. size of the extra dimension becomes singular the 
expansion scalar remains 
finite as long as $\dot b$ is finite. Therefore we expect $b(\eta)$ to play a role only in introducing a scaling effect. This will become clearer
in the next section, when we consider the general scenario where all 
the three non--constant metric functions are considered.

One may ask--what about analytic expressions for the kinematic variables
in the general case? 
Analytic expressions for the first integrals of the geodesic equations 
with non--constant $f(\sigma)$ and $a(\eta)$ can be found easily. 
However, the ESR as obtained from the definitions (Eq.\ref{eq:thetadef}-\ref{eq:omegadef}) are functions of $\sigma$,
$\eta$ and cannot be reduced to explicit functions of $\lambda$ 
alone. This happens because we do not know how $\sigma(\lambda)$ and 
$\eta(\lambda)$ are related analytically. Thus, the above two subcases 
are the only ones where one can find useful, closed--form analytic 
expressions for the ESR, directly from 
the velocity field.

\subsection{Numerical solutions}

Here, we numerically solve Eq.\ref{eq:gen} simultaneously with the geodesic 
equation for different combinations of the metric functions, in order to 
understand the interplay amongst all the terms appearing in the 
Raychaudhuri equations \cite{wald,toolkit}. The effect of the shear and 
rotation on the expansion are mutually opposite. Thus, the rotation may play 
a role in avoiding/delaying congruence 
singularities.  On the other hand, the curvature term 
$R_{AB}u^Au^B$ also has a significant effect on ESR profiles through its 
large positive or negative value at a given spacetime point. 

\begin{figure}[h!]
\subfigure[\,With $f(\sigma) = \log(\cosh\sigma)$, $a(\eta) = 2\eta$ and $b(\eta) = 1 + 1/\eta$]{\label{fig:comp_case1a}
\includegraphics[width = 2.6 in]{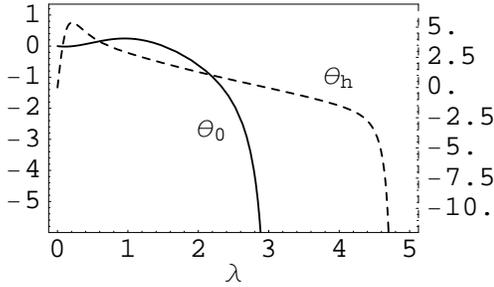}} \hspace{0.25 in}
\subfigure[\,With $f(\sigma) = -\log(\cosh\sigma)$, $a(\eta) = 2\eta$ and $b(\eta) = 1 + 1/\eta$]{\label{fig:comp_case2a}
\includegraphics[width = 2.5 in]{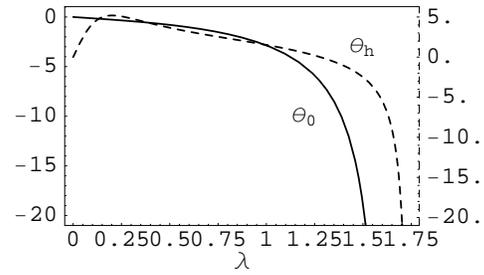}}

\subfigure[\,With $f(\sigma) = \log(\cosh\sigma)$, $a(\eta) = 1/(1-\eta)$ and $b(\eta) = 1 - \eta/2$]{\label{fig:comp_case3a}
\includegraphics[width = 2.65 in]{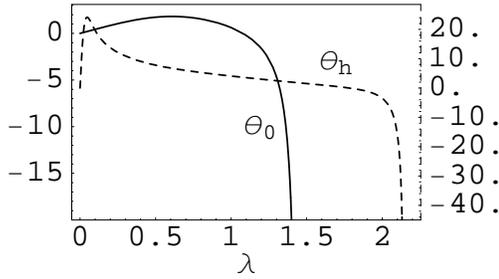}}\hspace{0.25 in}
\subfigure[\,With $f(\sigma) = -\log(\cosh\sigma)$, $a(\eta) = 1/(1-\eta)$ and $b(\eta) = 1 - \eta/2$]{\label{fig:comp_case4a}
\includegraphics[width = 2.5 in]{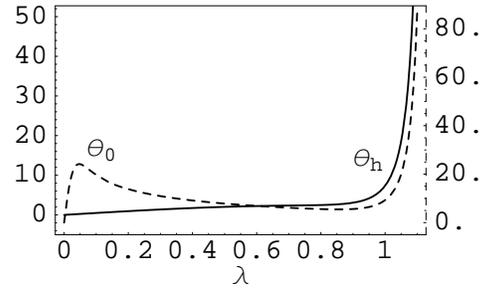}}

\caption{Nature of expansion scalar for different cosmological braneworld scenarios with two different sets of initial conditions. The subscripts $0$ ($h$) correspond to zero (high) initial rotation.} \label{fig:numerical}
\end{figure}

We have analysed each case for two types of initial conditions -- one with zero initial rotation and one with very high initial rotation, keeping initial $\Theta$ and $\Sigma_{AB}$ as zero. Non-zero initial values for $\Omega_{AB}$ are chosen such that $\Omega_{AB}u^B = 0 = u^A\Omega_{AB}$ at  $\lambda = 0$. Initial velocities for cases involving Set(A) scale factors are taken as $\{0.728, 0.1, 0.1, 0.1, 0.5\}$, whereas for Set(B), it is assumed as $\{1.1314, 0.1, 0.1, 0.1, 0.5\}$ so that the timelike constraint is satisfied. 
In Fig.\ref{fig:numerical}, we present the evolution of the expansion scalar for the different scenarios mentioned in Table \ref{tab:models}.


In the presence of a growing warp factor, a FRW (radiation dominated) brane and an asymptotically static extra dimension, geodesic congruences without any initial rotation, expand slowly at first but later 
become focused at a finite $\lambda$, as observed in Fig.\ref{fig:comp_case1a}. The shear ($\Sigma^2$) is found to
grow indefinitely (figure not shown). 
When high initial rotation is introduced, the congruence diverges 
for a very short 
while but eventually the geodesics get focused again at another finite but 
larger value of $\lambda$. Though rotation increases at late times, 
it is always dominated by shear, which grows even faster. Initial rotation 
only succeeds in delaying the focusing. The curvature term, as we note, 
is not an important factor here.

In contrast, Fig.\ref{fig:comp_case2a} shows that, in the presence of a decaying warp factor, a 
FRW (radiation dominated) brane and an asymptotically static extra dimension, the geodesics, 
without any initial rotation, do come closer to each other monotonically, 
but focus only at $\sigma\rightarrow\infty$. As before, the shear grows indefinitely. When high 
initial rotation is introduced, the congruence expands initially but eventually geodesics tend to focus again asymptotically. The curvature term becomes large positive valued and thus always seems to help in the occurrence of congruence singularities.

Fig.\ref{fig:comp_case3a} represents almost the same behavior as seen in Fig.\ref{fig:comp_case1a}. However, here the large initial rotation decays down very quickly. The physical reason behind this is the very fast expansion of spacetime that smears out all the initial rotation components.

On the other hand, in the presence of a decaying warp factor, 
a de Sitter brane with an asymptotically static extra dimension, in Fig.\ref{fig:comp_case4a}, we have an example where 
geodesics are defocused irrespective of the initial 
rotation. Even though it seems that at late times shear totally dominates over
rotation, it is the curvature term in the Raychaudhuri equation that becomes 
dominant and causes the defocusing. Remember that, with $a(\eta) = b(\eta) =$ constant, congruence singularity was inevitable. On the other hand, we have observed that initially the curvature term is very small. This implies that, in this case, a high enough initial negative expansion should lead to a congruence singularity at a finite $\lambda$ (before the curvature term becomes dominant). This behavior has been checked with an initial expansion, $\Theta = -30$ (figure not shown).

From the results in Fig.\ref{fig:numerical}, one can 
draw some general conclusions about the nature of the ESR variables. 
Congruence singularity is inevitable (irrespective of the initial rotation) in the cases addressed in 
Fig.\ref{fig:comp_case1a} and Fig.\ref{fig:comp_case3a}. 
This is because, in 
the presence of growing warp factor, the geodesics have a turning point in 
the extra dimension. This forces the congruence singularity to occur. 
Fig.\ref{fig:comp_case2a} represents a case where the geodesics are not bounded. 
Even in this case, a high initial expansion cannot make the congruence to diverge. 
On the other hand, the geodesics are divergent 
when $f(\sigma) = -\log(\cosh\sigma)$, $a(\eta) = 1/(1-\eta)$ and 
$b(\eta) = 1 - \eta/2$ [Fig.\ref{fig:comp_case4a}], which corresponds to a 
negatively warped and exponentially expanding brane. Here, it is interesting to 
note that the defocusing occurs because of the curvature term which becomes dominant and large negative, and has nothing to do with the initial rotation. 
Therefore, one can say that this defocusing is purely an effect of the spacetime geometry.

\begin{figure}[h!]
\subfigure[\,Evolution of 2D surface elements for $f(\sigma) = \log(\cosh\sigma)$, $a(\eta) = 2\eta$ and $b(\eta) = 1 + 1/\eta$ with $\Theta = \Sigma^2 = 0$ and $\Omega^2 \sim 54$ at $\lambda = 0$.]{\label{fig:fullsurf1_14}\includegraphics[scale=0.8]{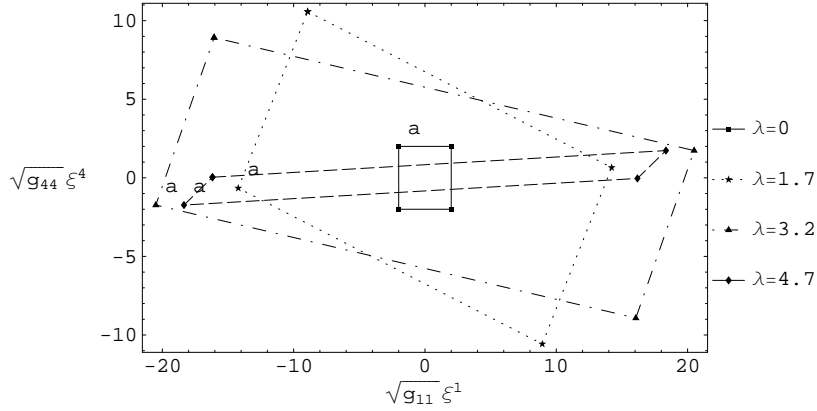}}
\subfigure[\,Evolution of 2D surface elements for $f(\sigma) = -\log(\cosh\sigma)$, $a(\eta) = 1/(1-\eta)$ and $b(\eta) = 1 - \eta/2$ with $\Theta = \Sigma^2 = 0$ and $\Omega^2 \sim 1080$ at $\lambda = 0$.]{\label{fig:fullsurf2_14}\includegraphics[scale=0.9]{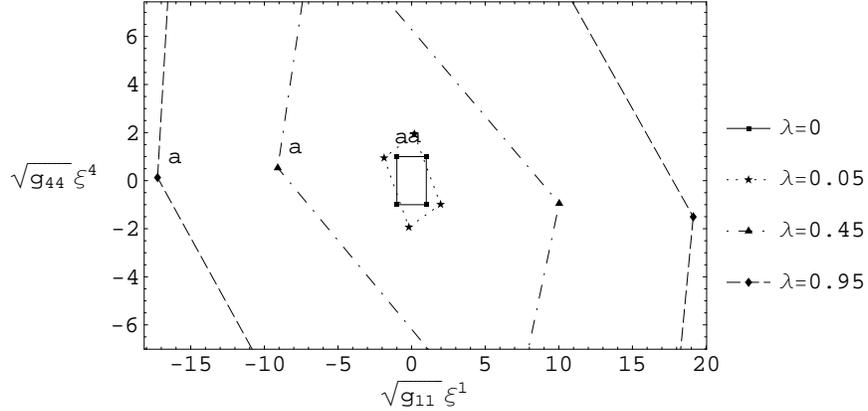}}
\caption{Evolution of the projected square elements on $\sqrt{g_{11}}\xi^1$-$\sqrt{g_{44}}\xi^4$ plane.} \label{fig:fullsurf1}
\end{figure}
Let us now look at the evolution of a congruence of geodesics from the 
local observer's point of view and see how the ESR profiles, 
plotted in Fig. \ref{fig:numerical}, are realised. 
As done earlier, here also we have plotted the evolution of square elements, 
orthogonal to the congruence of the timelike geodesics, projected on 
$\sqrt{g_{11}}\xi^1$-$\sqrt{g_{44}}\xi^4$ planes. These plots provide a different perspective of the evolution  
since they involve different shear and rotation tensor components. We have labeled one 
point of each area element (the one in the second quadrant, initially) 
as ``$\text{a}$''.  Following the location of this labeled point 
on future quadrilaterals (with increasing $\lambda$) one can see the 
effect of rotation. 

Fig.\ref{fig:fullsurf1_14} corresponds to the evolution of the dashed curve in Fig.\ref{fig:comp_case1a} (i.e. in presence of a growing warp factor). 
In Fig.\ref{fig:fullsurf1_14}, shrinking of the area element 
after an initial expansion, increase in the amount of shear after a decrease 
in the middle and a quick decrease in rotation are more clearly visible. 

Fig.\ref{fig:fullsurf2_14} represents the evolution of the dashed curve in Fig.\ref{fig:comp_case4a} (i.e. in presence of a decaying warp factor). The area of the quadrilateral keeps on increasing with $\lambda$. Effect of high initial rotation is prominent, so is its rapid decrease. 
Effect of shear in Fig.\ref{fig:fullsurf2_14} matches with the profile of Fig.\ref{fig:comp_case4a}. 
As mentioned earlier, these qualitatively different ESR evolutions 
are in one to one correspondence with different models. Therefore,
loosely speaking, the pictorial visualisation provides  
pointers toward possible verification of those models, though much more
needs to be done in order to arrive at explicit verifiable signatures.

\begin{table}[h]
\begin{center}
\begin{tabular}{|c|c|c|c|}
\hline
Warp factor & Constant $a(\eta)$, $b(\eta)$&
$a(\eta)=2\eta$, $b(\eta)=1+1/\eta$& $a(\eta)=1/(1-\eta)$,
$b(\eta)=1-\eta/2$ \\
$e^{2f(\sigma)}$& (Analytical results) & (FRW Radiation dominated) & (de Sitter) \\
\hline
Growing & Congruence singularity &  Congruence singularity &  Congruence singularity\\
        & at finite $\sigma$  & at finite $\sigma$ [Fig.\ref{fig:comp_case1a}]& at finite $\sigma$  [Fig.\ref{fig:comp_case3a}]      \\
\hline
Decaying & Congruence singularity &  Congruence singularity  & Defocusing at $\sigma\rightarrow\infty$ [Fig.\ref{fig:comp_case4a}]\\
        & at  $\sigma\rightarrow\infty$   &  at  $\sigma\rightarrow\infty$ [Fig.\ref{fig:comp_case2a}] & or congruence singularity at \\
        &                         &                          & finite $\sigma$ for large, -ve initial $\Theta$ \\
\hline
Constant & -- & No congruence singularity & No congruence singularity \\
\hline
\end{tabular}
\end{center}
\caption{Summary of behaviour of geodesic congruences for different metric coefficients}
\label{summary-table}
\end{table}

\section{Discussion}

We began by considering geodesic flows in the RS  background. 
Without branes congruence singularities always occur whereas 
with two branes, the expansion 
profile indicates how focusing/defocusing can occur in the spacetime 
between the branes. 

Later, we analyse geodesic flows in a
bulk geometry with a thick brane. 
Using first integrals of geodesic motion, analytic expressions for the 
kinematic variables are obtained. We
show how differences arise as we
change the warp factor from growing to decaying or when we do not have
any warping but retain the time-evolving cosmological and extra dimensional
scales. 

Further, we numerically solve the Raychaudhuri
and geodesic equations to obtain the 
expansion, shear, rotation and demonstrate the role of initial 
conditions on their evolution. With $a(\eta)=\eta$,
a growing warp factor leads to a finite $\eta$ 
(and $\sigma$) congruence 
singularity whereas for a decaying warp factor, geodesics 
are focused at $\eta$ (and $\sigma$) $\rightarrow \infty$.
For a de Sitter universe, 
a decaying warp factor may fail to focus the geodesics 
(due to the large negativity of the
curvature term in the Raychaudhuri equation), though this is not the case 
with a growing warp factor.   
When the curvature effect is relatively small, a congruence singularity 
can still arise but for high enough negative initial expansion.
The effect of initial rotation, on the ESR profiles, 
especially the expansion scalar, is found to be quantitative. 
Focusing 
without and with initial rotation yields 
similar results, though with large initial rotation, 
geodesics tend to spread for a while
(focusing at larger $\lambda$ value). All our conclusions are summarised 
in Table \ref{summary-table}.

For a visual perspective, we have shown
snap-shots of the evolution of a square element
orthogonal to a geodesic congruence, from a local observer's point of view. 
The evolution of the  expansion, shear and rotation,
along the congruences become more explicit through these figures.

The effect of $b(\eta)$ 
seems to be largely quantitative since, in our models, as 
$\lambda$ evolves $b(\eta)$ 
tends to a static value with a deceleration. 

It may be asked--what relevance, if any, does a congruence singularity have
in the context of realistic scenarios? After all, congruence singularities
are not real spacetime singularities where curvatures diverge. 
Here, we are tempted to draw an analogy from null geodesic congruences,
for which congruence singularities are nothing but the well-studied caustics
where optical intensities get magnified immensely. Similarly, in the case 
of timelike
geodesics, we may end up with accretion--like effects resulting out of matter
accumulation in the neighborhood of a point. For instance, our visualisation
analyses
do show how the square elements change shape, get rotated because of
variations in the metric functions. We may contemplate such accretion
effects for flows around brane--world black holes \cite{Pun:2008ua} and in such situations,
it will become necessary to pursue a line of thought very similar to
what we have followed in this article.

\section*{Acknowledgments}

SG thanks Council for Scientific \& Industrial Research (CSIR), India for 
providing financial support and Centre for Theoretical Studies, IIT Kharagpur, India for allowing him to use its research facilities.


\end{document}